\documentclass[apj]{emulateapj}
\newcommand {\Lya}    {Ly$\alpha$}   %  Lyalpha
\newcommand {\Ha}     {H$\alpha$}
\newcommand {\Hb}     {H$\beta$}
\newcommand {\HI}     {\ion{H}{1}}   %  HI
\newcommand {\CIV}    {\ion{C}{4}}

\newcommand {\kms}    {km~s$^{-1}$}
\newcommand {\NHI}    {$N_{\rm HI}$}

\newcommand {\tnma}{\tablenotemark{a}}
\newcommand {\tnmb}{\tablenotemark{b}}
\newcommand {\tnmc}{\tablenotemark{c}}
\newcommand {\tnmd}{\tablenotemark{d}}

\newcommand {\FUSE}  {{\it FUSE}} 

\newcommand {\etal}  {et~al.}

\slugcomment{ApJ 2011, submitted, v2}

\begin{document}

\title{Broad Ly$\alpha$ Emission from Three Nearby BL\,Lacertae Objects}

\author{John T. Stocke, Charles W. Danforth}
\affil{CASA, Department of Astrophysical and Planetary Sciences, University of Colorado, 389-UCB, Boulder, CO 80309; danforth@casa.colorado.edu, stocke@casa.colorado.edu}
\author{\& Eric S. Perlman}
\affil{Florida Institute of Technology, Physics and Space Sciences Department, 150 West University Blvd., Melbourne, FL 32901; eperlman@fit.edu}

%*****************************************************************************
%	ABSTRACT:
\begin{abstract}

We present far-UV HST/COS spectra of four nearby BL\,Lac Objects.
BL\,Lac spectra are dominated by a smooth, power-law continuum which
arises in a relativistic jet.  However, the spectra are not
necessarily featureless; weak, broad- and/or narrow-line emission is
sometimes seen in high-quality optical spectra.  We present detections
of \Lya\ emission in HST/COS spectra of Mrk\,421 ($z=0.030$) and
PKS\,2005$-$489 ($z=0.071$) as well as an archival HST/GHRS
observation of Mrk\,501 ($z=0.0337$).  Archival HST/STIS observations
of PKS\,2155$-$304 ($z=0.116$) show no \Lya\ emission to a very low
upper limit.  Using the assumption that the broad-line region (BLR)
clouds are symmetrically placed around the AGN, we use these measured
\Lya\ emission features to constrain either the relativistic $\Gamma$
values for the ionizing continuum produced by the jet (in the
ionization-bounded case) or the mass of warm gas (in the
density-bounded case). While realistic $\Gamma$ values can be obtained
for all four cases, the values for Mrk\,421 and PKS\,2155$-$304 are
high enough to suggest that covering factors of broad-line-region
clouds of $\sim$1--2\% might be required to provide consistency with
earlier values of Doppler boosting and viewing angles suggested for
this class of BL Lacs. This discrepancy also exists in the case of
M\,87, where the amount of Doppler boosting in our direction is
expected to be minimal, again suggestive of a small covering factor of
broad-line-region clouds.  If, as these small covering factors might
suggest, the assumptions of a density-bounded model could be more
correct, then the observed Lya\ luminosities require that BL\,Lac/FR~I
nuclei possess very little warm gas (10$^{-4}$ to 10$^{-5}$ M$_\odot$)
as suggested by Guilbert, Fabian \& McCray (1983).  If these clouds
are in pressure balance with a hotter ($\sim10^6$~K) gas, the BLR
contains too little mass to power the AGN by accretion alone.
\end{abstract}


%%%%%%%%%%%%%%%%%%%%%%%%%%%%%%%%%%%%%%%%%%%%%%%%%%%%%%%%%%%%%%%%%%
\section{Introduction} % long version of intro moved to old_intro.tex 9/14/10

BL\,Lacertae Objects (BL\,Lacs hereafter) are an extreme type of AGN
in which the non-thermal continuum emission is thought to be
relativistically-beamed and Doppler-boosted, overwhelming thermal
sources of emission seen in most other AGN classes.  These sources,
together with the flat-spectrum radio quasars, are the most numerous
and most luminous AGN at photon energies greater than a few hundred
keV \citep[e.g.,][]{Abdo09}.  While there are a few BL\,Lacs that
completely lack optical emission lines\citep[e.g.,][]{RectorStocke01,
Sbarufatti06}, high signal-to-noise ratio (SNR) optical spectroscopy
often detects weak but moderately luminous emission lines, especially
in objects with spectral energy distributions (SEDs) that peak in the
infrared \citep[``low frequency peaked BL\,Lacs'' or LBLs;
e.g.,][]{Stickel93,ScarpaFalomo97,RectorStocke01,Corbett00} like
BL\,Lac itself \citep{Vermeulen95}.

%% Target Table
 \begin{deluxetable*}{llllrl}
 \tabletypesize{\footnotesize}
 \tablecolumns{6} 
 \tablewidth{0pt} 
 \tablecaption{Observation Summary}
 \tablehead{\colhead{BL\,Lac Object}   &
            \colhead{RA (J2000) Dec} &
            \colhead{$z_{\rm AGN}$}    &
            \colhead{Instrument}     &
            \colhead{$t_{\rm exp}$} &
	    \colhead{Obs. Date}    }
 	\startdata
	Mrk\,421        & 11:04:27 $+$38:12:32 & 0.0300 & COS/G130M &  1738 & 2009, Dec 24 \\
	                &                      &        & COS/G160M &  2404 & 2009, Dec 24 \\
	PKS\,2005$-$489 & 20:09:25 $-$48:49:54 & 0.0710 & COS/G130M &  2462 & 2009, Sept 21 \\
	                &                      &        & COS/G160M &  1854 & 2009, Sept 21 \\
	Mrk\,501        & 16:53:52 $+$39:45:37 & 0.0337 & GHRS/G160M& 29,196& 1992, Feb 26-28\\
	PKS\,2155$-$304 & 21:58:52 $-$30:13:32 & 0.116  & STIS/E140M& 14,244& 1999, Nov 9 \\
	                &                      &        &           & 14,244& 2000, Sept 26 \\
 	\enddata
 \end{deluxetable*}

Less is known about the emission line regions of the high-energy
peaked BL\,Lacs \citep[HBLs;][]{PadovaniGiommi95}, which have SEDs
peaking in the UV to X-rays and extended radio emission consistent
with being \citet{FanaroffRiley74} type 1 (FR\,1s)
\citep{PerlmanStocke93,RectorStocke01,Giroletti08}.  Most HBLs have no
detectable optical line emission \citep {RectorStocke01,Sbarufatti06}.
It is possible that emission lines comparable to FR\,1s (weak
\Ha$+$[\ion{N}{2}] and/or weak [\ion{O}{2}]) are present in many HBLs
but are hidden at optical wavelengths by the nonthermal AGN continuum
and by starlight from the host galaxy \citep{BrowneMarcha93}.  On the
other hand, many HBLs possess optical spectra which contain the
absorption lines and edges (e.g., \ion{Ca}{2} H\&K break, G-band)
typical of giant elliptical galaxies that are the usual hosts of these
AGN
\citep{HutchingsNeff92,Wurtz96,Scarpa00,Sbarufatti05}. 

While the LBL/HBL distinction may be a result of previous selection
techniques \citep[as suggested by][]{Collinge05}, the observed
properties of these two classes do vary; all HBLs and most LBLs have
radio emission consistent with being beamed FR\,1s \citep{Rector00}
while some LBLs are consistent with beamed emission from the more
powerful Fanaroff-Riley class 2 (FR\,2s) sources like the
flat-spectrum radio quasars \citep[e.g.,][]{Brinkmann96,
RectorStocke01}.  Setting aside the few FR\,2-like LBLs, most BL\,Lacs
can be explained as beamed FR\,1 radio galaxies.  It is unclear
whether the differences between LBL and HBL are a consequence of
viewing angle \citep{PerlmanStocke93,Jannuzi94,RectorStocke01,
Nieppola08}, peak frequencies and luminosities \citep[the so-called
"Blazar sequence"; see][]{Fossati98,Ghisellini98,Ghisellini02,
CavaliereDelia02,Guetta04,Maraschi08,GhiselliniTavecchio08}, or a
combination of both geometric and intrinsic factors
\citep{GeorganopoulosMarscher98}.

Broad-line emission is an essential property of AGNs and can provide
insights on a number of properties.  Fundamentally, the line
luminosities represent a direct measure of the amount of circumnuclear
gas at parsec scales and the isotropic, ionizing photon environment.
Direct measurement of the luminosity of the broad-line region (BLR) in
BL\,Lacs can provide a new, direct measurement of the radiative power
of these AGN and so can provide an independent constraint on the
amount of beaming.  Current evidence suggests that the luminosity of
any broad emission lines in BL\,Lacs may be linked to the synchrotron
peak frequency as originally suggested by \citet{GeorganopoulosMarscher98} and
\citet{Ghisellini98}.  If BLR photons are abundant, the cooling of the jet's
synchrotron emission is enhanced greatly by the increased
inverse-Comptonization of these photons by the jet.  This increased
cooling will decrease the peak frequency of the jet emission,
resulting in a peak at IR energies or below (i.e., LBLs).  However,
without BLR seed photons, the jet cooling will be dominated by the
synchrotron process and so is much less efficient.  This results in a
much higher-frequency synchrotron peak, in the UV or X-rays (i.e.,
HBLs).

Is the lack of observed line emission in BL\,Lacs, particularly HBLs,
an intrinsic property of these less-luminous AGN, or is it a product
of the photon environment induced by the jet?  The creation of stable
BLR clouds may be inhibited by soft X-ray spectra \citep{Guilbert83}
which are observed in only one of the two BL\,Lac classes.  In the
case of LBLs, the X-ray emission is relatively weak and the X-ray
spectrum is hard \citep[$\alpha_x<1$;][]{Urry96,Padovani04}.  In HBLs
the X-rays are quite luminous ($L_X=10^{44-46}\rm~ergs~s^{-1}$) and
soft \citep[power-law spectral index in energy $=-1$ to $-2.8$;]
[]{Perlman96,Perlman05,Padovani01}.  Thus, the soft X-ray continuum
seen in HBL-type objects may be impeding their ability to create BLR
clouds.  This can be confirmed by observational data using
\Lya, the most sensitive probe of BLR gas.

In this Letter we present far-ultraviolet (FUV) spectroscopy of four
of the nearest HBL prototypes, in which weak \Lya\ emission lines have
been detected or sensitive upper limits can be set.  In Section 2, we
present new high SNR spectra from the Cosmic Origins Spectrograph
(COS), newly-installed on the Hubble Space Telescope (HST) of Mrk\,421
and PKS\,2005$-$489, an archival Goddard High Resolution Spectrograph
(GHRS) spectrum of Mrk\,501, and an archival Space Telescope Imaging
Spectrograph (STIS) spectrum of PKS\,2155$-$304.  We measure \Lya\
line widths and luminosities and reddening-corrected continuum
luminosities and slopes for these HBLs.

These observations provide the first simultaneous measurements of
continuum and BLR line emission in HBLs.  In Section~3, we use the
observed continua to derive ionizing luminosities and thereby predict
the \Lya\ line luminosities under the ``nebular hypothesis''
(ionization-bounded case).  Comparison between the predicted and
observed \Lya\ luminosity allows an estimate of the beaming angle for
the ionizing flux.  Using the density-bounded assumption allows us to
compute the amount of warm BLR gas assuming that the gas is
optically-thin.

%********************************************************************************
%% Full COS spectra of Mrk421 and PKS2005-489
\begin{figure*}[t]
  \epsscale{.9}\plotone{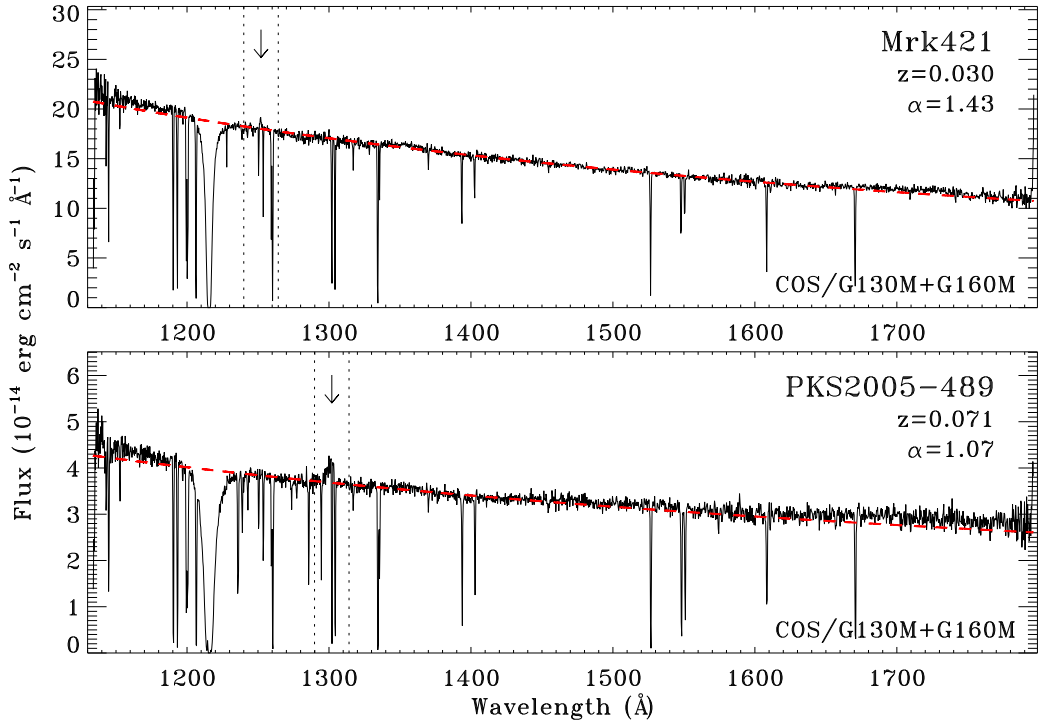} %%% fig_cosdata.pro %%%%%%%%%
  \caption{COS observations of Mrk\,421 (top) and PKS\,2005$-$489
  (bottom) over the entire COS/G130M$+$G160M range.  The flux has been
  corrected for reddening and are shown binned by 30 pixels ($\sim4$
  resolution elements).  Power law continua of the form
  $F(\lambda)=I_0*(\lambda/912)^{-\alpha_\lambda}$ were fitted to
  line-free continuum regions and are shown as dashed curves.  The
  narrow absorption features are mostly interstellar absorption lines;
  a few lines blueward of \Lya\ are intergalactic.  The region
  around the rest-frame \Lya\ (arrow) bounded by vertical dotted lines
  is expanded in Figure~2 below.}  \label{fig:fullcos}
\end{figure*}

\section{BL\,Lac Observations and Analysis}

We present new COS FUV spectra of two BL\,Lac targets (Mrk\,421 and
PKS\,2005$-$489) as well as archival observations of Mrk\,501 (GHRS)
and PKS\,2155$-$304 (STIS).  The four targets are summarized in
Table~1.

%% COS obs

COS far-UV observations of Mrk\,421 and PKS\,2005$-$489 were carried
out during the first three months of COS science observations as part
of the COS Guaranteed Time Observations (PID 11520, PI Green).  Four
exposures were made in each of the G130M ($1135<\lambda <1480$~\AA)
and G160M ($1400<\lambda<1795$~\AA) medium-resolution gratings
($R\approx18,000$) for each target.  Four central wavelength settings
at each grating dithered known instrumental features along the
spectrum and provided continuous spectral coverage over
$1135<\lambda<1795$~\AA\ \citep[see][]{Green10,Osterman10}.

All COS exposures were reduced using {\sc CalCOS v2.11f}.
Flat-fielding, alignment, and coaddition of the processed exposures
were carried out using IDL routines developed by the COS GTO team
specifically for COS FUV data\footnote{See {\tt
http://casa.colorado.edu/$\sim$danforth/costools.html} for our
coaddition and flat-fielding algorithm and additional discussion.}.
The details of the coaddition method are discussed in
\citet{Danforth10}.  Briefly, each exposure was corrected for narrow,
$\sim15\%$-opaque shadows from repellor grid wires.  The local
exposure time in these regions was reduced to give them less weight in
the exposure-weighted final coaddition.  Similarly, exposure times for
data at the edges of the detectors was de-weighted.  With four
different central wavelength settings per grating, any residual
instrumental artifacts from grid-wire shadows and detector boundaries
have negligible effect on the final spectrum.

Strong ISM absorption features in each exposure were aligned via
cross-correlation and interpolated onto a common wavelength scale.
The wavelength shifts were typically on the order of a resolution
element ($\sim0.07$~\AA, $\sim17$~\kms) or less.  The coadded flux at
each wavelength was taken to be the exposure-weighted mean of flux in
each exposure.  To quantify the quality of the combined data, we
identify line-free continuum regions at various wavelengths, smooth
the data by the seven-pixel resolution element, and define $S/N
(\equiv \rm mean(flux)/stddev(flux)) \approx 37$ and $\approx 20$ for
the Mrk\,421 and PKS\,2005$-$489 observations, respectively.  The
fully-reduced, coadded spectra of Mrk\,421 and PKS\,2005$-$489 are
shown in Figure~\ref{fig:fullcos}.

%% Mrk501 and PKS2155 details

In addition to the two COS datasets, we analyze archival observations
of the BL\,Lac objects Mrk\,501 (GHRS) and PKS\,2155$-$304 (STIS).
Mrk\,501 was observed for 29 kiloseconds with the G160M grating
covering $1222<\lambda<1257$~\AA\ at a FWHM of $\sim20$ \kms\ and
SNR$\approx15$.  PKS\,2155$-$304 was observed for 28.5 ksec with the
STIS/E140M grating ($1140<\lambda<1729$~\AA), FWHM$\sim7$~\kms,
SNR$\approx27$.  Details of the GHRS and STIS datasets and reduction
techniques can be found in \citet{Penton00} and
\citet{IndebetouwShull04}, respectively.

%% dereddening, N(HI), etc

In order to accurately measure BL\,Lac continua, the observations were
corrected for Galactic extinction.  Traditional measures of extinction
are based on broad-beam radio observations of Galactic \HI\ emission
and may not take into account small-scale variations in this value.
Since three out of four of our datasets cover the region around
Galactic \Lya\ absorption, we can directly measure \NHI\ and calculate
the extinction, $E(B-V)$, along our sight line via the relationship
${\rm E(B-V)}=N_{\rm HI}/5.8\times10^{21}~\rm cm^{-2}$
\citep{ShullvanSteenberg85}.  Uncertainties in fitted Galactic $N_{\rm
HI}$ and the resulting errors in extinction correction dominate the
subsequent uncertainties in continuum fitting\footnote{In Figure~1,
the best-fit reddened power-laws show some systematic deviation at
both the long and short waveleght ends for unknown reasons--a single
power law index may not be applicable, the COS flux calibration by not
be correct, or the far-UV extinction correction may be deficient.
These discrepencies do not affect our measurements of \Lya\ emission
although the continuum flux at $\lambda<912$ \AA\ may be underestimated
by $\sim10$\%.}  The Mrk\,501 observations do not include 1216\AA, so
we use the literature extinction value for this sight line.  Observed
spectra were corrected for Galactic reddening via the
\citet{Fitzpatrick99} parameterization.

% continuum fitting

Dereddened spectra were next blueshifted into the rest frame of the
BL\,Lac and line-free continuum regions were identified and fitted
with a power law of the form $F_\lambda=I_0\,
(\lambda/912)^{-\alpha_\lambda}$.  The region near rest-frame \Lya\
was not included in the fits.  Flux at the rest-frame Lyman continuum
is derived via a short extrapolation from the HST spectra.  The
Mrk\,501 GHRS data has such a small wavelength coverage that the
continuum fit slope and LyC extrapolation are uncertain.

%% Lya emission

%% Detail of Lya emission features
\begin{figure*}[t]
  \epsscale{.95}\plotone{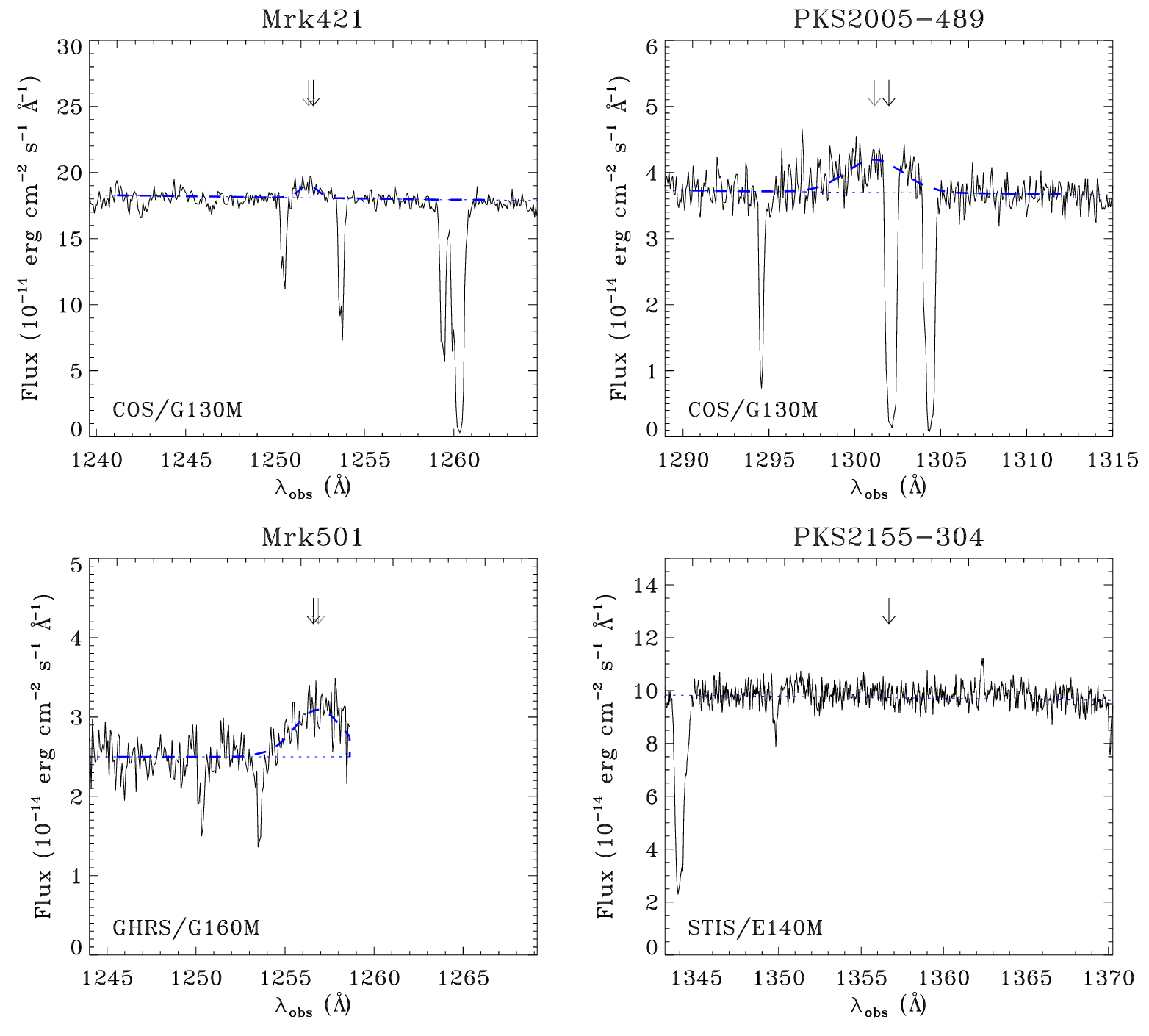} %%% fit_cont.pro %%%%%%%%%%%%%
  \caption{Detailed spectra of the \Lya\ region in each BL\,Lac.  All
  data are dereddened, binned to $\sim15-20$ \kms, and shown over the
  range $cz_{\rm AGN}=\pm3000$ \kms.  Power-law continuum fits are
  shown as dotted lines.  Black arrows mark the systemic velocity of
  \Lya\ using the absorption-line redshifts for these objects.  Thick
  dashed curves show Gaussian \Lya\ fits for the three detections and
  line centroid locations are marked with gray arrows.  Full fit
  parameters are given in Table~2.  All narrow absorption features are
  intervening interstellar or intergalactic absorption features
  unrelated to the AGN emission.}  \label{fig:Lyafits}
\end{figure*}

Next, the \Lya\ emission feature was measured.  All three observed
emission features were well-fit with single Gaussian profiles with
free parameters $v_{\rm centroid}$, FWHM, and integrated intensity
$I({\rm Ly\alpha})$.  Total \Lya\ luminosity (assuming isotropy) is
then $L({\rm Ly\alpha})= I({\rm Ly\alpha})\,4\pi(c\,z_{\rm em}/H_0)^2$
with $H_0=72\rm~km~s^{-1}~Mpc^{-1}$.  The significance of these
features can be estimated as
\begin{equation}
SL\approx\sqrt{C}\,\frac{W\,SN_{\rm res}}{\sqrt{FWHM}}
\end{equation}
where the scaling constant $C$ is equal to the number of resolution
elements per \AA.  This comes to $\sim9\sigma$, $\sim15\sigma$, and
$\sim23\sigma$ for Mrk\,421, PKS\,2005$-$489, and Mrk\,501,
respectively.

No \Lya\ emission is seen in the spectrum of PKS\,2155$-$304.  If we
assume FWHM$\sim3$~\AA\ for the emission line width we find a
$4\sigma$ upper limit of $W\la44$~m\AA\ or $I(\rm Ly\alpha)\la
4\times10^{-15}\rm~erg~cm^{-2}~s^{-1}$.  Given the higher continuum
luminosity of PKS\,2155$+$304, the \Lya\ luminosity upper limit
($L(\rm Ly\alpha)\la1.1\times10^{41}~erg~s^{-1}$) is comparable the
detections in the other HBLs.

All of the measured and derived quantities discussed above are listed
in Table~2 for each of the four BL\,Lacs.  

%% Quantities Table
% last update 10/20/10
\begin{deluxetable*}{lccccl}
 \tabletypesize{\footnotesize}
 \tablecolumns{6} 
 \tablewidth{0pt} 
 \tablecaption{BL\,Lac Measured and Derived Quantities}
 \tablehead{\colhead{Quantity} &
            \colhead{Mrk\,421} &
	    \colhead{PKS\,2005$-$489}&
	    \colhead{Mrk\,501} &
            \colhead{PKS\,2155$-$304} &	
            \colhead{Units}    }
  \startdata
   $z_{\rm em}$                 & 0.030     & 0.071       & 0.03366    & 0.116  &              \\
   $\log\,N(\rm HI)$            &$20.04\pm0.03$ &$20.49\pm0.03$ & 20.04:\tnma & $19.96\pm0.03$& \\
   E(B-V)                       &$0.019\pm0.002$ &$0.053\pm0.004$ & 0.019:\tnma & $0.016\pm0.001$& \\
   $\rm S/N_{res}$              & 37        & 20          & 15         & 27     & (see text) \\
  \cutinhead{Continuum Fit: $F_\lambda=I_0\,(\lambda/912)^{-\alpha_\lambda}$} 
   $\alpha_\lambda$             &$1.43\pm0.02$&$1.07\pm0.04$&$ 0.7: $&$ 1.13\pm0.01$&  \\
   $I_0    $                    &$27.2\pm0.6$&$5.01\pm0.22$&$ 3.0:  $&$ 13.4\pm0.2$& $\rm (10^{-14}~erg~cm^{-2}~s^{-1}~\AA^{-1})$\\
  \cutinhead{X-ray Spectral Fit Parameters\tnmb}
   $a$                          &$2.10-2.58 $ &$ 2.36-3.14 $  &$ 1.41-2.18$  &$ 2.3-2.8$& \\
   $b$                          &$0.31-0.46$  &$0.27\pm0.09$&$ 0.12-0.56 $ & $-0.3-0.65$ & \\
  \cutinhead{Ly$\alpha$ Emission Feature}		
   $\lambda_{\rm centroid}$     &1251.9     & 1301.1      & 1256.9     & 1357\tnmc& (\AA, observed)  \\
   $v_{\rm Ly\alpha}-v_{\rm em}$&$-60$      & $-190$      & $+70$     & \nodata & (\kms)  \\
   FWHM                         &$1.23\pm0.13$&$4.25\pm0.24$&$3.31\pm0.33$& 3\tnmc & (\AA) \\
   FWHM                         &$300\pm30  $&$1050\pm60$&$820\pm80  $& 740\tnmc& (\kms)  \\
   Equivalent Width             &$-76\pm7$   & $-467\pm47$& $-830\pm83$     & $<44$  & (m\AA) \\
   Signficance Level            &9          & 15          & 23         & $<4$   & ($\sigma$) \\
   $I(Ly\alpha)$                &$1.27\pm0.12$&$2.38\pm0.11$&$2.2\pm0.2$&$\la0.4$& $(\rm 10^{-14}~erg~cm^{-2}~s^{-1})$ \\
   $L(Ly\alpha)$                &$2.37\pm0.22$&$24.9\pm1.1$&$5.2\pm0.3$&$\la11$ & $(\rm 10^{40}~erg~s^{-1})$ \\
  \cutinhead{Predicted \Lya\ Emission}
   $I(Ly\alpha)\rm_{predicted}$&$3.0\times10^{-10}$&$3.6\times10^{-11}$&$\sim10^{-11}$&$1.0\times10^{-10}$& $(\rm erg~cm^{-2}~s^{-1})$ \\
   Overprediction factor (OPF)  &$2.4\times10^4$           & $1.5\times10^3$&$\sim600$&$>2.6\times10^4$ & \\
   Doppler Factor $(\delta)$\tnmd  & 16.6    &   6.5 &  $\sim4$ & $>13.6$ & \\
   Minimum Required $(\Gamma)$\tnmd&  8.3    &   3.2 &  $\sim2$ & $>6.8$  & \\
   Maximum Viewing Angle\tnmd  &  3.5    &   8.8 & $\sim16$ & $<4$    & (degrees) \\
  \enddata
 \tablenotetext{a} {Literature value; not directly measured.}
 \tablenotetext{b}{X-ray spectral fit parameters as defined in \citet{Perlman05}: $dN/dE\propto E^{[-a+b\,\log(E)]}$.  Parameter ranges are taken from the literature of the last ten years \citep{Acciari09,Brinkmann01,Brinkmann03,Brinkmann05,Donato05,Edelson01,Foschini06,Massaro08,Padovani01,Ravasio04,Sembay02,Tramacere07a,Tramacere07b,Zhang06a,Zhang06b,Zhang08}.}
 \tablenotetext{c}{Assumed quantity; see text.} 
 \tablenotetext{d}{Assuming unity covering factor of BLR clouds}
\end{deluxetable*}

% CIV emission upper limits paragraph 2/15/11
Neither Mrk\,421 nor PKS\,2005$-$489 show emission from the \ion{C}{4}
$\lambda\lambda$ 1548, 1550 doublet.  Due to the lower effective area
of COS at the red end of the far-UV range, these data are of lower S/N
than those covering the Lya emission region.  We derive upper limits
on \ion{C}{4} emission for these two objects of $I_{\rm CIV}\la
0.8\times10^{-14}~\rm erg~cm^{-2}~s^{-1}$ and $I_{\rm CIV}\la
4.3\times10^{-14}~\rm erg~cm^{-2}~s^{-1}$ for Mrk\,421 and
PKS\,2005$-$489, respectively.  The spectra of the other two BL\,Lac
objects do not cover the rest-frame $\lambda\sim1549$ \AA\ region.

\section{Inferences for the Broad-line Regions of BL\,Lacs}

With both continuum and emission-line measurements, we can estimate
the covering fraction and beaming angle for BL\,Lacs under the nebular
hypothesis.  To start we assume that the covering factor of the BLR
clouds is 100\% and the clouds are spherically-distributed around the
nucleus. From this simple geometry other geometries can be assumed, so
that these calculations can be used as baseline values.  First, we
estimate the total ionizing flux in the BLR by extrapolating the
power-law continuum to higher energies.  Next, we predict the
resulting BLR \Lya\ line emission assuming that a) the ionizing flux
is isotropic (clearly a poor assumption for BL\,Lacs) and, b) the
broad line clouds are optically thick to ionizing photons with unity
covering factor (i.e., the ``nebular hypothesis'' or
ionization-bounded case) OR, c). the BLR clouds are optically-thin
individually and in toto (the density-bounded case). We will then
interpret the difference between the predicted and observed \Lya\
luminosities in the ionization-bounded case to determine if the
required Doppler boosting is consistent with estimates from other
considerations. And then we use the \Lya\ line luminosities to infer
BLR warm gas mass in the density-bounded case.

\subsection{Ionization-bounded Model}

Under Case-B recombination, the number of photons emitted in a given
line (e.g., \Hb) can be related to the total number of ionizing
photons by the ratio of their recombination coefficients \citep[see
Eq. 5.40 of][]{OsterbrockFerland06}, 
\begin{equation}
\frac{L(H\beta)}{h\,\nu_{H\beta}}\approx \biggl(\frac{\alpha^{eff}_{H\beta}}{\alpha_B}\biggr)\,\int_{\nu_0}^{\infty}{\frac{L(\nu)}{h\,\nu}\,d\nu}.
\end{equation}
where the effective
$\alpha^{eff}_{H\beta}=3.03\times10^{-14}\rm~cm^{-3}~s^{-1}$ and
$\alpha_B=2.59\times10^{-13}\rm~cm^{-3}~s^{-1}$ for Case-B
recombination at $T\sim10^4$~K.  \Lya\ luminosity $L(Ly\alpha)$ is
then related to $L(H\beta)$ by the ratio of their respective specific
intensities, 
\begin{equation}
\frac{j_{Ly\alpha}}{j_{H\beta}}=\biggl(\frac{h\,\nu_{Ly\alpha}}{h\,\nu_{H\beta}}\biggr)\,\biggl(\frac{\alpha_B}{\alpha^{eff}_{H\beta}}\biggr)\approx34
\end{equation}
under Case-B recombination.  Substituting the power law fitted to the
observed BL\,Lac spectra, Eq.~2 becomes
\begin{equation}
I(H\beta)_{\rm pre.}= h\,\nu_{H\beta}\,\biggl(\frac{\alpha^{eff}_{H\beta}}{\alpha_B}\biggr)\,\frac{I_0\,\lambda_0^2}{c}\,\int_{\nu_0}^{\infty}\frac{1}{h\,\nu}\,\biggl(\frac{\nu}{\nu_0}\biggr)^{-(2-\alpha_\lambda)}d\nu
\end{equation}
and the predicted \Lya\ flux is
\begin{equation}
I(Ly\alpha)_{\rm pre.}=4\,\biggl(\frac{\alpha_B}{\alpha^{eff}_{H\beta}}\biggr)\,I(H\beta)_{\rm pre.}
\end{equation}

%% optically-thick case B

The \Lya\ emission predicted from the power-law continuum fit
over-predicts the observed \Lya\ emission by 2--4 orders of magnitude
in all four cases (Table~2). The simplest interpretation of the \Lya\
over-prediction is that the ionizing continuum seen by the BLR clouds
is 2--4 orders of magnitude less intense than what we see (i.e.,
beaming).  In this case we can use the over-prediction factor ($OPF$)
to calculate a required beaming factor ($\delta =
[\Gamma(1-\beta\,cos\theta)]^{-1}$, where the relativistic $\Gamma$
and $\beta$ are as usually defined and $\theta$ is the ``viewing
angle'' between the outflow axis and the observer's line-of-sight),
the minimum required $\Gamma$ of the outflow and the maximum allowed
viewing angle all assuming a unity covering factor of BLR clouds as
seen from the source.  Since the bulk of the ionization of the BLR
clouds is due to photons just blueward of the Lyman Limit, the
$OPF=\delta^{3+\alpha_{\nu}}$, where the power-law spectral index in
frequency $\alpha_{\nu}=(\alpha_{\lambda}+2)$ (see Table~2).  See
\citet{UrryPadovani95} Appendices~A \&~B for equations and derivations.  
The exponent on $\delta$ is appropriate for a spherical cloud
morphology of the emitting region; a continuous, cylindrical
morphology yields one less power on this exponent and thus more
extreme values of $\Gamma$ than we quote in Table~2. Also, these
expressions may not be exact in these BL\,Lacs because we do not
directly observe the flux and spectrum of the ionizing radiation as a
function of all off-axis angles; we are assuming that the dominant
off-axis ionizing radiation source is the relativistic jet.

On the other hand, for the large majority of viewing angles, the
dominant UV/X-ray photon source may be the nuclear region of the AGN,
not the jet.  BL\,Lacs and their likely parent population of FR\,1
galaxies have little or no disk emission suggesting a very low
accretion rate in a radiatively-inefficient accretion flow
\citep{Allen06} onto their central black holes.  As discussed in 
\citet{Allen06}, such a flow will also have a soft X-ray spectrum,
albeit one that is thermal in shape, with a typical temperature of
0.5--1 keV.  If the \Lya\ BLR emission we observe in these BL Lacs is
due to thermal accretion emission then the value for the OPF only sets
lower limits on $\Gamma$ and $\delta$ since the off-axis jet emission
must be even less luminous than what we have assumed for the values in
Table~2.  Either this thermal spectrum or the off-axis jet spectrum
(which peaks in the far-UV -- soft X-ray with a soft X-ray spectrum,
as described in e.g., \cite{Perlman05}) would be hostile to the
formation of BLR clouds (see below).

While the values of $\delta$, minimum $\Gamma$, and maximum viewing
angle in Table~2 are reasonable in the context of some HBLs
\citep{Hovatta09,Ghisellini10}, $\Gamma = 3-5$ and $\theta=10^\circ$
are suggested from other considerations
\citep{UrryPadovani95,Giroletti06}; $\Gamma=3$ is suggested from
comparisons in luminosity functions of HBLs and FR\,1 radio galaxies
\citep{UrryPadovani95} and $\Gamma=3-5$ is suggested from radio 
core-to-extended flux ratios \citep{PerlmanStocke93,Rector00,
Giroletti06}.  Therefore the $\Gamma$ values in Table~2 for Mrk\,421
and PKS\,2155$-$304 are somewhat too high to be consistent with
earlier estimates. If we use the best previous values for $\Gamma$ and
$\theta$ quoted above to determine $\delta$, we can then use the
observed OPF to obtain a best estimate of the BLR cloud covering
factor. In this case values of 1--2\% are found.  Such low covering
factors are much more the expectation of the density-bounded
model. However, it is equally possible that the values of $\Gamma$ in
the Mrk\,421 and PKS\,2155-304 jets are significantly higher than
estimates from entire samples of HBLs and so are consistent with an
ionization bounded model. Indeed, by assuming that the
ionization-bounded model is correct, the values of $\Gamma$ in Table 2
are the first such estimates using this method.

%%% Reality check with Seyfert *** use Haardt/Madau rad field ***

As a conceptual check of our methodology, we perform the equivalent
analysis on a representative unbeamed AGN with strong \Lya\ emission.
Seyfert galaxies lack the relativistic jet of radio-loud AGN and the
anisotropy of their emission is presumed to be low.  Recent HST/COS
observations of the nearby Sy\,1.5 galaxy Mrk\,817 show strong \Lya\
emission ($I(\rm Ly\alpha)=8.1\times10^{-12}\rm~erg~cm^{-2}~s^{-1}$)
and a power-law continuum with parameters $I_0=1.5\times10^{-13}
\rm~erg~cm^{-2}~s^{-1}$ and $\alpha_\lambda=1.3$ \citep{Winter10}. 
We extrapolate the AGN continuum via the Haardt \& Madau (2005)
spectrum for Akn\,120 (another nearby Seyfert galaxy) and calculate
the total ionizing flux as above. The predicted \Lya\ line emission is
$2.7\times10^{-11}\rm~erg~cm^{-2}~s^{-1}$, a factor of only $\sim3$
higher than observed. Given that a Lyman limit break is not observed
in this class of objects, a covering factor of $f_{\rm BLR}\sim30\%$
is reasonable.

A second more relevant comparison for our inferences is the case of
M87, the FR\,1 radio galaxy Virgo~A.  \citet{Sankrit99} presented an
HST/Faint Object Spectrograph (FOS) spectrum of the nucleus of M87
arguing that the detected \Lya\ emission had to be primarily nuclear,
photionized emission by comparison to a second spectrum taken through
an aperture displaced 0.6 arcsec nearby.  The \Lya\ emission of M87 is
at low enough redshift that it is significantly affected by Galactic
\Lya\ absorption, whose influence can only be estimated very
roughly.  \citet{Sankrit99} estimate that M87's intrinsic \Lya\ is at
least a factor of two larger than the $9\times10^{-14}$ ergs cm$^{-2}$
that they measure.  If we use double their value for the \Lya\ flux
and take the observed continuum flux extrapolated to 912\AA\ using a
$\nu^{-1}$ power-law typical of the 4 BL Lacs presented here, a value
of $OPF=100-200$ is obtained.  The \citet{Sankrit99} observations are
consistent with this spectral index but have a wavelength range too
small to determine an accurate $\alpha_{\nu}$.  This large OPF is
intriguing given the observed spectral properties of M87 and its
inferred substantial off-axis viewing angle \citep[$25^{\circ}$;][]
{Heinz97}.  The minimum values of $\Gamma$ required by the large
jet-to-counterjet ratio for the M87 jet ($3-5$) suggest that the
continuum is not substantially beamed in our direction ($\delta \sim$
1.5--2 or less if $\Gamma$ is larger) so that the OPF should be quite
low, 5--10 times lower than what we find. While other modeling of
apparent superluminal motions within the jet \citep{biretta99}
suggests slightly smaller viewing angles ($\sim19^{\circ}$), somewhat
larger values for $\Gamma$ are required by these motions, again
suggesting minimal beaming in our direction.  Additionally, some of
the nuclear \Lya\ emission could be shock-heated \citep{Dopita97}
which would further reduce the amount attributable to the ionizing
continuum. As with our more extreme BL Lac Objects the high OPF for
M87 suggests a low covering factor of a few percent for the BLR
clouds. Since in the case of M87 there is less chance for
misinterpretation due to potentially large beaming factors, these low
covering factors are again suggestive of the density-bounded case.
 
\subsection{Density-bounded Model} 

A second physical interpretation is that the \Lya\ overprediction is
not due to beaming but to the physical conditions in and distribution
of the BLR clouds; i.e., the clouds themselves are optically-thin and
may also have small covering factor so that most of the ionizing
photons simply escape regardless of the beaming. The 1--2\% covering
factors found by applying best values to the ionization-bounded case
may be indicative that the density-bounded case is more correct for
this class of objects (HBLs and FR\,1 radio galaxies). Up to now we
have assumed a large-scale-isotropic distribution of optically-thick
clouds; however they may be optically thin and/or anisotropic.  The
BLR clouds may be more highly concentrated in the equatorial region of
the AGN and sparser at the poles \citep[e.g.,][]{Elvis00}.  Regardless
of their geometry, the emission line diagnostic tells us only about
the BLR clouds irradiated by the continuum we observe and that we
assume is coming from the jet.

All four of our objects are observed to have steep X-ray spectra.  The
early work of \citet{Guilbert83} points out that a soft X-ray source
heats circum-nuclear gas to a single high temperature phase
$T\sim~10^6$~K, avoiding the creation of a stable two-phase medium and
so inhibits the formation of large warm clouds in the BLR. Most AGN
X-ray spectra are uniformly hard with spectral index $\alpha~\sim~
-0.7$ compared to the much steeper spectra seen in these four (see
Table~2) and most other HBLs.

%% LyC decrement
Clearly the neutral gas column {\it within the beam} is low enough
that no Lyman continuum decrement is seen.  All four targets in this
study have \FUSE\ spectra covering the rest-frame Lyman limit and we
see no significant change in continuum level on either side of this
break.  While it is not a strong constraint on the BLR neutral column,
we can say $N_{\rm HI}\la\rm few\times10^{16}~\rm cm^{-2}$ for three
of our four targets, on average within the beam.  (The \FUSE\ data for
Mrk\,501 is of poor quality and we can only set a limit $N_{\rm
HI}\la1.5\times10^{17}\rm~cm^{-2}$ for this object.)  This may be due
to low optical depth, small covering fraction, or a combination of
these effects, but in any case supports a BL\,Lac BLR model which is
optically-thin.

In the optically-thin regime, we can estimate the amount of warm gas
in the BLR by assuming that there is one hydrogen atom for every \Lya\
photon emitted.  The observed \Lya\ luminosities for these objects
(Table~2) are so low that this assumption yields a BLR with only
$\sim10^{-4}$ to $10^{-5}~M_\odot$ of warm gas.  Larger amounts of gas
could be present if the few \Lya\ emitting clouds in the BLR are
themselves optically thick at the Lyman limit so that most of the mass
is not emitting any \Lya.  This cloud geometry is however at variance
with simple models of advection-dominated or Bondi-Hoyle accretion
flows \citep[ADAF;][]{Narayan95,Allen06}.  Assuming pressure
equilibrium between the \Lya\ emitting clouds ($T\sim10^{4.5}$~K) and
a dominant coronal phase ($T\sim10^6$~K) suggested by the Guilbert,
Fabian \& McCray scenario, and assuming a BLR size of a few parsecs,
yields a mass of only $\sim 0.1~M_\odot$.  The amounts inferred to
power the observed AGN \citep[$\dot
m\la10^{-2}~M_\odot~yr^{-1}$]{CavaliereDelia02} are however much
larger than the amounts that can be supplied by the BLR gas we infer
to be present in these objects.  There appears to be so little
circumnuclear gas in BL\,Lacs that it is unclear how these sources
power their relativistic jets.

%%EP Addition 

It is worth considering the fact that these lines are not very broad
at all, having $FWHM\la1000$ \kms (\citet{Sankrit99} measure a weak,
broad component to \Lya\ in M87 that could have $FWHM\la3000$ \kms,
although the continuum is poorly defined in their FOS spectrum).  By
comparison, typical broad lines in Seyfert galaxies have FWHM many
times higher, and the same is also true of the broad lines found in
LBLs \citep{Corbett00,ScarpaFalomo97, Vermeulen95}.  In fact the line
widths we observe, are much closer to those seen in narrow, forbidden
lines such as [\ion{O}{3}].  Therefore, in addition to the
possibilities presented above, the \Lya\ emitting material could be
further out in these objects, at radii $>10$~pc, more typical of NLR
gas.  In this case there maybe even less gas available for accretion
power in the BLR.

\section{Conclusions}

We have measured weak, broad \Lya\ emission in three nearby
high-energy peaked BL\,Lac objects (HBLs) and set a sensitive upper
limit for a fourth.  Our \Lya\ luminosities are
$\sim10^{41}\rm~erg~s^{-1}$ in three objects, an order of magnitude
stronger than the weak, narrow optical emission lines that have been
observed in a few HBLs \citep[$\log\,L\leq40\rm~ergs~s^{-1}$;][]
{Rector00,Sbarufatti06}.  To the best of our knowledge this is the
first detection of UV line emission from this rare class of AGN
\citep[although \Lya\ emission was observed in M87 by ][]{Sankrit99}.
BL\,Lacs are an important probe of physics at the very highest
energies and our simultaneous measurements of both continuum and line
emission will help constrain the models of their structure and
kinematics.

From the observed data, we make a simple estimate of the number of
ionizing photons produced by the jet and predict the \Lya\ line
emission if this entire energy were radiated isotropically into
optically thick broad-line-region clouds.  The predicted \Lya\
emission is 2--4 orders of magnitude larger than what is observed and
we interpret this over-prediction in two possible ways:\\

\noindent {\bf (1)} The over-prediction is a symptom of relativistic 
beaming angle and broad line cloud covering fraction.  For our two
most extreme cases (Mrk\,421 and PKS\,2155$-$304), the beaming angles
are somewhat smaller than predicted via other means and we needed to
invoke a 1--2\% covering factor to be consistent with earlier beaming
estimates for HBLs \citep{UrryPadovani95}. However, it is equally
plausible that the relativistic $\Gamma$ for these two sources are
much higher ($\sim8\times$) than the HBL population as a whole. But in
the case of M87, for which \Lya\ and a weak UV continuum were detected
using FOS \citep{Sankrit99}, the over-prediction factor of
OPF$\geq$100 is intriguing because little beaming is expected in our
direction. In this case a larger value of $\Gamma$ does not explain
the large OPF value since the large viewing angle means that the
continuum we observe is largely unbeamed.  In this case a small
covering factor is the most likely cause of the weakness of the
\Lya\ emission in M87.  

\noindent {\bf (2)} Small covering factors inferred for M87 and 
possibly for Mrk\,421 and PKS\,2155-304 are quite similar to having
optically-thin and/or sparse clouds. So a quite plausible
interpretation of the paucity of \Lya\ photons coming from these
BL\,Lacs is that their BLRs are density-bounded, not
ionization-bounded.  This result was predicted for BL\,Lac objects by
Guilbert, Fabian \& McCray in 1983, as a consequence of their steep
X-ray continua.

Conclusion (2) can be extended in two ways.  First, if the BLR clouds
are optically thin, not all of the ionizing radiation is captured and
converted to line emission and the \Lya\ luminosities are set by the
total mass in warm gas in the BLR, $10^{-4}$ to $10^{-5}\rm~M_\odot$.
Assuming pressure balance in the BLR between these warm clouds and a
hot medium as in the Guilbert et al. model, requires that there is
only $\sim$ 0.1 M$_\odot$ of gas in the BLR in toto. This is far too
little to power the AGN that we observe in these objects. Therefore,
other fueling mechanisms need to be considered for BL\,Lac Objects and
their parent population, FR\,1 radio galaxies.  One obvious way around
this conclusion is that, while there are very few BLR clouds around
the nucleus in BL Lacs and FR\,1 radio galaxies, these clouds are each
very optically-thick, so that a considerable amount of mass can be
hidden by being shielded from the ionizing continuum.  However,
optically-thick clouds seems at variance with an advection-dominated
accretion-flow (ADAF; \citet{Narayan95}) scenario for fueling these
sources.

Weak, broad \Lya\ line emission has been detected in the nucleus of
the FR\,1 radio galaxy M87 \citep{Sankrit99} with HST/FOS at
comparable line luminosity to the Mrk\,421 detection reported here.
However, the low resolution (which blends the emission with damped
Galactic \Lya\ absorption) and signal-to-noise (which poorly detects
the UV continuum) is insufficient to determine the BLR parameters for
M87 conclusively. There is also the issue that near-nuclear emission
is thought to be due to shocks like in the spectra of LINERS
\citep{Dopita97} so that the photo-ionized \Lya\ could be even weaker
than what is observed.  Assuming the best available values for
$\Gamma$ and viewing angle in the case of M87 requires small covering
factors in the ionization-bounded solution as in the cases of Mrk\,421
and PKS\,2155$-$304. Thus, as with the four BL Lacs, a density-bounded
scenario should be considered for these low-luminosity AGN which
leaves the question of their power source still open. The nucleus of
M87 will be observed in HST Cycle 18 with COS obtaining much bettter
resolution and signal-to-noise than the
\citet{Sankrit99} spectrum.
 
In addition to the above, one must now wonder about the differences
between the LBL and HBL classes of BL\,Lacs.  As already pointed out,
a few LBLs have considerably more luminous broad-line emission than
seen in these HBLs and other LBLs \citep[e.g.,][]{Stickel93,
ScarpaFalomo97,RectorStocke01,Corbett00,Vermeulen95}, more consistent
with the nebular hypothesis.  In a purely viewing angle unification of
LBLs and HBLs only a very tiny number of BLR clouds would be
irradiated by a hard X-ray continuum (the LBL ``region'' of the
celestial sphere as seen from the nucleus) while more BLR clouds would
see the soft X-ray continuum characteristic of HBLs. In the few LBLs
with luminous broad emission lines it appears likely that much larger
amounts of gas are present in the BLR as quasi-stable broad-line
clouds.  If this is the case the off-axis ionizing continuum must be
hard enough to permit BLR cloud formation, at variance with the simple
viewing angle unification scheme outlined above.  And also, given the
much larger gas mass in the BLR, standard accretion can be a viable
power source for these FR\,2-like LBLs. These differences argue
against a purely viewing angle based unification of all LBLs and HBLs.

%% are these still BLLacs?
%It is a myth that spectra of BL\,Lac objects are uniformly featureless; indeed many show weak line emission in the optical \citep[others?]{ScarpaFalomo97}.  Typically, a threshhold of $W<5$~\AA\ in the optical is used to define the class \citep[e.g.,][]{Stickel93} and, ironically, BL\,Lac itself has occasionally violated this definition \citep{Vermeulen95}.  Scaling the equivalent width criterion to ultraviolet wavelengths, we note that all four of our objects show $W<1$~\AA.  Furthermore, we observe line widths $FWHM<1000$~\kms; broad, but not as broad as seen in other BL\,Lac objects.  {\bf [more here?]}

%% variability discussion?
%{\it [Discussion of low/high state of HBLs.  PKS\,2005$-$489 was observed in a low state; the continuum flux is roughly a factor of ten lower than expected.  Does this mean jet emission is being quenched by IC photons from the BLR?  What about the other objects in the sample?  I haven't been able to find much about these yet.]}

% Future work

Additional work is needed to flesh out the relationship between the
two BL\,Lac classes, as well as the UV line-emission properties of
BL\,Lacs in general.  Mrk\,421 and PKS\,2005$-$489 were observed early
in Cycle~17 as part of the COS Guaranteed Time Observation (GTO)
program.  Two higher-redshift BL\,Lacs (1ES\,1028$+$511 and
PMNJ\,1103$-$2329) are scheduled for observation during Cycle 18 and
we will analyze their \Lya\ emission properties.  Two additional
BL\,Lacs at unknown redshift (S5\,0716$+$714 and 1ES\,1553$+$113) will
also be observed and we will look closely for emission features by
which the systemic redshift can be determined.  The latter object was
observed briefly in 2009 and the redshift was constrained via
intervening \Lya\ absorbers \citep[$z>0.4$;][]{Danforth10}.  A
reobservation of M87 with HST/COS and STIS can be used to determine if
the \Lya\ emission in FR\,1s is consistent with expectations of
unified schemes. Detection of both \Lya\ and \CIV\ emission can test
whether the gas is ionized by shocks or by a UV continuum. The spatial
resolution along the STIS slit will determine how much of the observed
\Lya\ and continuum flux is truly nuclear.

%% detectability of lines
%Issues of inverse-Compton seed photons aside, it is worth noting that the detectability of broad emission lines actually {\it increases} with continuum flux.  The significance level of a line detection is proportional to the equivalent width of the line and the S/N of the data (Eq.~1).  For a line of constant equivalent width, the contrast is goes as $F_{\rm cont}^{-1}$, but significance level goes as $F_{\rm cont}^{1/2}$.

\medskip
%\medskip
%% Acknowledgements

It is our pleasure to acknowledge the many thousands of people who
made the HST Servicing Mission~4 the huge success that it was.  We
furthermore thank Steve Penton, St\'ephane Beland, and the other
members of the COS ERO and GTO teams for their work on initial data
calibration and verification and Lisa Winter for the Mrk\,817 fit
parameters.  JTS acknowledges a College Scholar Fellowship from the
College of Arts \& Sciences, U. of Colorado at Boulder for support
during this research work.  JTS also acknowledges the Center for
Computational Cosmology, Durham U., the Institute of Astronomy,
Cambridge U. and the Specola Vaticana, Castel Gandolfo for hospitality
during portions of this work.  CWD wishes to acknowledge a fruitful
discussion with members of the KIPAC consortium.  This work was
supported by NASA grants NNX08AC146 and NAS5-98043 to the University
of Colorado at Boulder.

\end{document}